\documentclass[%
  preprint,
  amsmath,amssymb, 
  aps, prfluids,
  superscriptaddress,
]{revtex4-2}

\usepackage{amsmath,amssymb}
\usepackage{stmaryrd}
\usepackage{graphicx}
\usepackage{dcolumn}
\usepackage{bm}
\usepackage{siunitx}
\usepackage[hidelinks]{hyperref}
\usepackage{placeins}

\newcommand{\maybeincludegraphics}[2][]{%
  \IfFileExists{#2}{\includegraphics[#1]{#2}}{%
    \fbox{\parbox[c][0.28\textheight][c]{\linewidth}{}}%
  }%
}

\begin{document}

\title{Deformation and instability of sessile soap bubbles in an electric field}

\author{Hongsik Kim}
\affiliation{Biological and Environmental Engineering, Cornell University, Ithaca, New York 14853, USA}
\email{hk866@cornell.edu}

\author{Sunghwan Jung}
\affiliation{Biological and Environmental Engineering, Cornell University, Ithaca, New York 14853, USA}
\email{sj737@cornell.edu}

\date{\today}

\begin{abstract}
Interfacial deformation under electric fields is a common phenomenon in many industrial processes. Particularly, we are interested in the dynamics of sessile soap bubbles in a parallel-plate electric field which exhibits a stable deformation regime followed by conical instability. Using side-view imaging, we track the equilibrium shapes, the transition to the unstable regime, and the pre-jet apex dynamics within one experimental system. In the stable regime, the meridional profile is well described by a spheroidal fit, and the aspect ratio collapses across initial bubble sizes onto a single steady-state branch when plotted against the dimensionless field $E^\ast = \sqrt{\mathrm{Bo}_e}$ for data acquired within a fixed ambient session where the electric Bond number $\mathrm{Bo}_e$ is defined as $\varepsilon_0 E_0^2 R_0/(2\gamma)$. The endpoint of this branch marks the transition to the unstable regime. Above onset of instability, the apex sharpens into a cone with half-angle $30.0^{\circ}$ $\pm$ $0.6^{\circ}$, below the classical Taylor value. To quantify the late pre-jet stage, we define the axial distance $b(t)$ from the instantaneous apex to a fixed reference vertex determined from the terminal cone geometry and measure its evolution. The corresponding rate grows as jetting is approached, and a near-tip inertia-capillary model captures the observed logarithmic trend as an approximation. Together, these measurements establish a single-system experimental benchmark in which stable electrocapillary deformation is organized by a single steady-state branch that leads into conical instability and pre-jet dynamics.
\end{abstract}


\maketitle

\section{Introduction}

Electric fields deform fluid interfaces through Maxwell stress acting against capillary pressure. The same balance underlies electrosprays, electrospinning, tip streaming, and cone-jet atomization, and it has been analyzed through classical electrohydrodynamics, reduced theories, and modern reviews of cone-jet operation~\cite{Zeleny1917,Taylor1964,Taylor1966,MelcherTaylor1969,Saville1997,DeLaMora2007,Vlahovska2019,GananCalvoEtAl2018}. These responses are also attractive because they enable contact-free control of charge transport and liquid delivery with high precision in electrospray ionization, particle synthesis, and thin-film deposition~\cite{CloupeauPrunetFoch1989,HayatiBaileyTadros1986,GomezTang1994,BarreroLoscertales2007,Jaworek2007,JaworekSobczyk2008}. For initially smooth interfaces, a common sequence is steady deformation, loss of equilibrium, and the formation of pointed tips, jets, or bursts~\cite{TorzaCoxMason1971,Sherwood1988,BasaranScriven1989,BasaranScriven1990,WohlhuterBasaran1992,VizikaSaville1992,LacHomsy2007}.

A classical reference point is Taylor's conical construction, which yields the well-known half-angle of $49.3^{\circ}$ for an equipotential cone in a compatible electrostatic field~\cite{Taylor1964}. That result, however, belongs to an idealized limit. Subsequent analyses have emphasized that cone-like critical shapes need not be unique and that field geometry, boundary conditions, and charge transport can modify the observed angle and the route to jetting~\cite{StoneListerBrenner1999,Yarin2001,Collins2008,Collins2013}. Related analyses of conical-point solutions for liquid-liquid interfaces and of surface-attached droplets in strong electric fields further show that selected cone angles and jetting pathways depend sensitively on electrical properties, geometry, and attachment conditions ~\cite{Ramos1994,Reznik2004}. In particular, Yarin \emph{et al.} argued that the Taylor cone does not represent a unique critical shape and reported experimentally observed half-angles closer to non-Taylor critical solutions in parallel capacitor-like geometries~\cite{Yarin2001}.

Compared with droplets, soap bubbles and soap films provide a thin-film, multi-interface platform in which electrohydrodynamic deformation can be studied in a comparatively simple geometry. Early experiments by Wilson and Taylor and by Macky documented bursting and deformation of soap bubbles in electric fields~\cite{WilsonTaylor1925,Macky1930}. Related bubble and film problems include electrically stressed gas cavities in liquids, charged hemispherical soap bubbles, soap-film bridges, and, more recently, deformation of sessile and floating soap bubbles in a plane capacitor~\cite{GartonKrasucki1964,HiltonVanderNet2009,MoultonPelesko2008,Mawet2021} and in non-uniform electrode configurations \cite{PeleszKapralZylka2023}. A related recent study also examined soap-bubble stream charging from analytical and experimental perspectives, extending the bubble-electrostatics literature beyond deformation alone ~\cite{Pelesz2020}. These systems make the departure from the canonical droplet problem explicit. The interface is a thin film with two air--liquid surfaces. The substrate fixes the large-scale geometry. The background electric field is set by a finite electrode gap rather than by a self-similar conical field.

Recent soap-bubble experiments in plane-capacitor geometries established field-driven deformation, but they left open how the steady deformation regime ends and how the apex evolves in the final approach to jetting~\cite{Mawet2021}. More broadly, the missing step is a unified experimental description that connects the stable branch, its endpoint, and the subsequent cone-forming and pre-jet stages within one soap-bubble system.

In this paper, we address that gap using a sessile soap bubble in a planar electric field. We first characterize the stable regime and ask whether the deformation can be organized by a single steady-state branch across initial bubble size when expressed in the dimensionless field $E^\ast = \mathrm{Bo}_e^{1/2}$ where $\mathrm{Bo}_e=\varepsilon_0 E_0^2 R_0/(2\gamma)$ compares electric stress with the capillary restoring stress of the soap film. We then identify the endpoint of that branch and track the subsequent conical sharpening up to jet emission. The main results are threefold. First, stable deformation is well represented by a spheroidal geometry and within a fixed ambient session the data collapse onto a single steady-state branch $\alpha(E^\ast)$. Second, the endpoint of that branch marks the transition to the unstable regime. Third, in the unstable regime the system selects cone angles of $30.0^\circ \pm 0.6^\circ$, and a fixed reference vertex allows comparing pre-jet trajectories across bubble sizes, with a near-tip inertia-capillary model capturing the observed logarithmic trend in $|db/dt|$.

\section{Experimental system and analysis methods}

\subsection{Experimental setup and electric-field protocol}

The apparatus consisted of a 3D-printed PLA insulating fixture carrying copper electrodes in a nominal parallel-plate arrangement. The insulation was shaped so that only the opposing electrode faces were exposed to the active region. The electrode gap was adjustable during preliminary tests. All data reported here were acquired at a fixed gap $H = \SI{50}{mm}$, chosen as a practical value that suppressed visible corona over the operating voltage range while still allowing the desired deformation and jetting responses. The lower electrode was grounded, and the high-voltage connection to the upper electrode was routed through a separate insulated compartment outside the imaging region to minimize perturbations of the field in the measurement zone.

\begin{figure*}[!t]
\centering

\begin{minipage}[t]{0.78\textwidth}
    \centering
    \maybeincludegraphics[width=\textwidth]{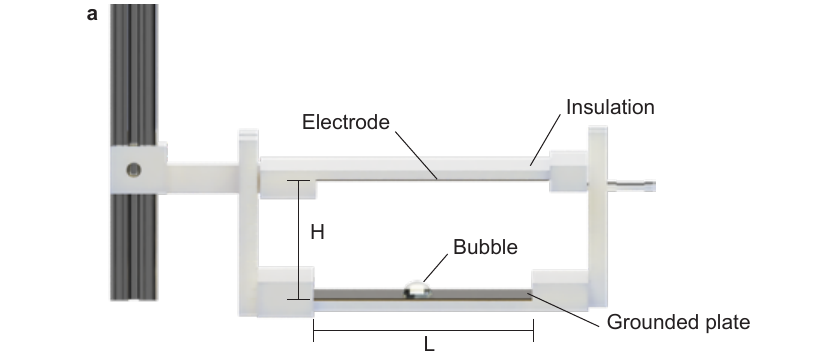}
\end{minipage}
\vfill
\begin{minipage}[t]{0.8\textwidth}
    \centering
    \maybeincludegraphics[width=\textwidth]{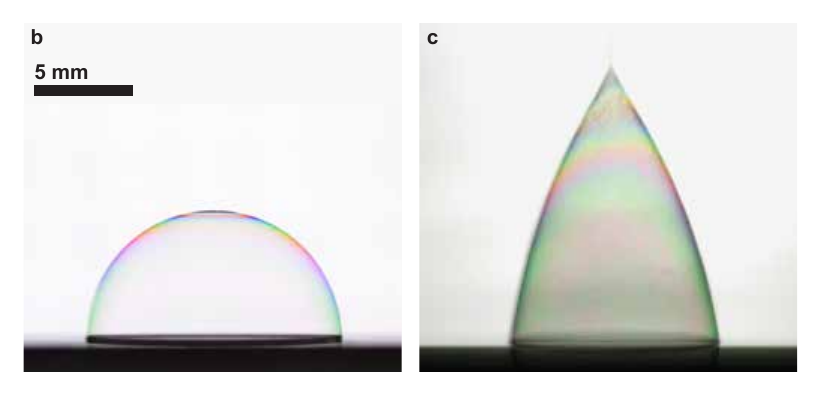}
\end{minipage}

\caption{\label{fig:fig1}
(a) Schematic of experimental setting for a nominally uniform electric field configuration. A sessile soap bubble of measured radius $R_0$ is pinned on the lower electrode and placed a distance $H$ below a top plate of width $L$. A nominal field of magnitude $E_0 \simeq V/H$ is applied and increased quasi-statically.
Representative side-view images illustrating the two response regimes: (b) smooth equilibrium deformation in the stable regime and (c) conical sharpening followed by jetting in the unstable regime.}
\end{figure*}

A sessile soap bubble of measured initial radius $R_0$ was formed on the lower electrode beneath a square copper top plate of characteristic lateral size $L\approx \SI{140}{mm}$ (Fig.~\ref{fig:fig1}). Because the plate is much larger than the bubble and the bubble was positioned near the plate center, the background field over the bubble-scale region is nominally uniform. We therefore use the nominal applied field
\begin{equation}
E_0 \equiv V/H,
\end{equation}
where $V$ is the applied potential difference. A Laplace calculation for the finite electrode geometry was used only to confirm that the field at the bubble location is well approximated by the parallel-plate estimate $V/H$. It was not used to determine the local field near the sharpened tip. A Laplace calculation for the exact electrode geometry is provided in Appendix C. Over the bubble-scale region beneath the plate center, the axial field differs from the parallel-plate estimate $V/H$ by less than about 0.02\%, which justifies the use of $E_0=V/H$ as the nominal control parameter for the stable regime scaling.

\begin{figure}[t]
\centering
\maybeincludegraphics[width=0.6\columnwidth]{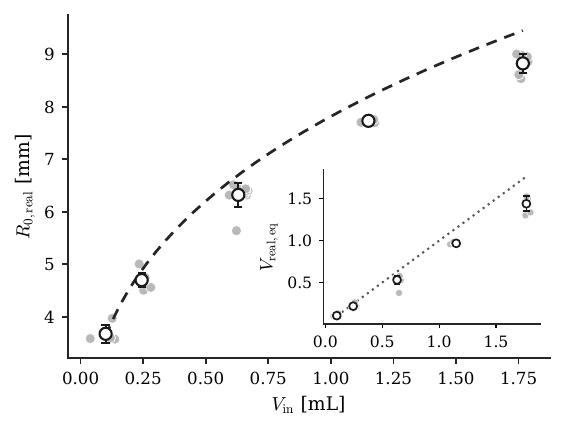}
\caption{\label{fig:fig_calibration} 
Calibration between pump input volume $V_{\mathrm{in}}$ and the measured initial bubble size. Gray symbols denote individual runs, and open circles with error bars indicate the mean $\pm$ standard deviation of the measured initial radius $R_{0,\mathrm{real}}$ for each input setting. The dashed curve is the hemispherical-equivalent radius associated with the pump input volume. Inset shows the hemispherical-equivalent volume estimated from the measured initial radius. The dotted line is the identity. All subsequent analysis uses the measured $R_{0,\mathrm{real}}$. 
}
\end{figure}

Before inflation, an excess \SI{2}{\micro\liter} sample of the soap solution was placed on the lower electrode to stabilize the contact region and reduce premature rupture. Air was then injected using a syringe pump with prescribed input volume $V_{\mathrm{in}}$ to form a sessile bubble, and only cases in which the bubble remained centered beneath the top plate were retained for analysis. Although bubbles were generated with prescribed $V_{\mathrm{in}}$, all analysis is based on the measured radius $R_0$ from images, because tubing compliance and bubble-formation dynamics can produce a mismatch between the pump setpoint and the measured bubble size. Fig.~\ref{fig:fig_calibration} reports the calibration between $V_{\mathrm{in}}$ and the measured $R_0$, together with the corresponding hemispherical-equivalent volume estimated from the measured radius. All subsequent nondimensionalization therefore uses the measured radius rather than the pump setpoint.

The voltage was increased quasi-statically in discrete steps. In the stable regime, images were acquired only after the interface had relaxed to a stationary shape, and the voltage increment was refined near the stability limit. Beyond a critical field, the apex no longer relaxed to a smooth equilibrium but instead continued to sharpen toward jetting. The principal data sets used for the collapse analysis were acquired within the same experimental session for each size group.

\subsection{Imaging and profile extraction}

Side-view images were used to extract the meridional profile. Stable-regime shapes were recorded with a Nikon D7100 camera. Unstable-regime dynamics were recorded with a Photron FASTCAM NOVA S9 high-speed camera equipped with an AF Micro-NIKKOR 105 mm lens at 9000 fps with a shutter time of \SI{5}{\micro\second}. High-speed sequences were recorded from the onset of visible deformation through jet emission and rupture. The present analysis uses the pre-rupture interval relevant to cone formation and late pre-jet apex dynamics. Before each acquisition, a ruler was imaged in the same optical configuration to establish the pixel-to-length conversion.

Customized MATLAB scripts were used to measure the overall bubble height and width from the calibrated images. Custom Python scripts were used to extract the meridional profile for subsequent fitting and model comparison. The image-processing workflow therefore consisted of spatial calibration, profile extraction, and geometric fitting.

In the stable regime, the extracted meridional profile is well described by a spheroidal profile with vertical and horizontal semi-axes $a$ and $b$, respectively, and the deformation is quantified by the aspect ratio
\begin{equation}
\alpha \equiv a/b.
\end{equation}
Here $\alpha=1$ corresponds to the undeformed reference state. The spheroidal fit is used as a geometric description, not an exact free-boundary solution.

In the unstable regime, the near-tip profile is fit locally by two straight segments defining a cone. The cone half-angle $\theta$ is measured between the cone side and the symmetry axis. To quantify the final pre-jet stage, we define a fixed reference vertex from the first frame with visible jet emission or, equivalently, from the last pre-jet frame in which the cone shapes are stationary. Specifically, local linear fits are applied to the near-tip cone shape in that reference frame, and the intersection of their extrapolated extensions defines the reference vertex. For each earlier frame, $b(t)$ is taken as the axial distance from the instantaneous apex to this fixed reference vertex. In high-speed videos, $t = 0$ denotes the first frame with visible jet emission.

\subsection{Solution properties and dimensionless groups}

Soap bubbles were prepared from an aqueous surfactant mixture of 63\% deionized water,
30\% glycerol, and 7\% DAWN commercial dish soap by volume. A representative
single-interface surface tension is taken as $\gamma \approx 0.03\,\mathrm{N/m}$, and the
liquid density is taken as $\rho \approx 1080\,\mathrm{kg/m^3}$.

The film thickness $h$ was not measured in situ during each run. We therefore use $h$
only as a literature-based order-of-magnitude estimate when checking the thin-film
condition $h \ll R_0$ and when setting the prefactor scale in the reduced apex-focusing
model. Reported equilibrium soap-film thicknesses range from common black films of
order $10$--$100\,\mathrm{nm}$ to Newton black films of order $5$--$10\,\mathrm{nm}$, while very
thin soap bubbles can approach limiting thicknesses of order $200\,\mathrm{nm}$ \cite{Evers1999,Chatzigiannakis2021}.
In view of this spread, we adopt $h = O(10^2\,\mathrm{nm})$ and use it only as a representative
range. The stable-regime collapse and the regime classification do not rely on a direct
measurement of $h$.

The reduced scaling used below is from the usual electrohydrodynamic
free-boundary formulation\cite{MelcherTaylor1969,Saville1997,Vlahovska2019}. In the gas region surrounding the bubble, the electric potential
satisfies
\begin{equation}
\nabla^2 \phi = 0,
\qquad
\mathbf{E} = -\nabla \phi,
\label{eq:laplace_phi}
\end{equation}
and the corresponding Maxwell stress tensor is
\begin{equation}
\mathbf{T}_M
=
\varepsilon_0
\left(
\mathbf{E}\otimes\mathbf{E}
-
\frac{1}{2}E^2\mathbf{I}
\right).
\label{eq:maxwell_tensor}
\end{equation}
For a single deformable interface, the local normal stress balance takes the form
\begin{equation}
\llbracket -p\mathbf{I} + \mathbf{T}_M \rrbracket \cdot \mathbf{n}
=
\gamma \kappa\,\mathbf{n},
\label{eq:normal_stress_balance}
\end{equation}
where $\kappa = 1/R_1 + 1/R_2$ is the total curvature and $\gamma$ is the surface tension of a single air--liquid interface \cite{MelcherTaylor1969,Saville1997,Vlahovska2019}. For the present soap bubble, the film has two air--liquid
interfaces, so the capillary restoring contribution enters the reduced scaling with an effective coefficient $2\gamma$ \cite{WilsonTaylor1925,Macky1930}. At the scaling level used here, the electric stress is therefore of order $\varepsilon_0 E_0^2$, whereas the restoring capillary stress is of order
$2\gamma/R_0$. This is the origin of the factor $2\gamma$ used below in the electric--capillary
control parameter. In the present experiments, $E_0 = V/H$ is used as the nominal
background field, as justified by the finite-electrode Laplace calculation in Appendix~C.

We therefore define the electric Bond number
\begin{equation}
Bo_e \equiv \frac{\varepsilon_0 E_0^2 R_0}{2\gamma},
\label{eq:Boe}
\end{equation}
and, for plotting convenience, its square root
\begin{equation}
E^* \equiv \sqrt{Bo_e}
=
E_0 \sqrt{\frac{\varepsilon_0 R_0}{2\gamma}}.
\label{eq:Estar}
\end{equation}
All collapse plots in the present work use the measured $R_0$ from images rather than the pump setpoint $V_{\mathrm{in}}$.

\begin{table}[t]
\caption{\label{tab:params}Representative measured and adopted system parameters.}
\begin{ruledtabular}
\begin{tabular}{l c l}
Quantity & Value (range) & Description \\
\hline
$R_0$ & $3$--$8\,\si{mm}$ & Measured initial bubble radius \\
$H$ & $\SI{50}{mm}$ & Electrode gap \\
$L$ & $\SI{140}{mm}$ & Characteristic top-electrode lateral size \\
$E_0$ & up to $10^5$--$10^6\,\si{V/m}$ & Nominal applied field magnitude \\
$\gamma$ & $\approx \SI{0.03}{N/m}$ & Single-interface surface tension; reduced capillary scale uses $2\gamma/R_0$ \\
$h$ & $h = O (10^2\,\mathrm{nm})$ & Estimated film thickness; not measured in situ \\
$\rho$ & $\approx \SI{1080}{kg/m^3}$ & Liquid density \\
$\varepsilon_0$ & $\SI{8.854e-12}{F/m}$ & Vacuum permittivity \\
\end{tabular}
\end{ruledtabular}
\end{table}

\section{STABLE DEFORMATION AND TRANSITION TO INSTABILITY}

\subsection{Spheroidal description of the steady profiles in the stable regime}

\begin{figure*}[t]
\centering
\maybeincludegraphics[width=0.8\textwidth]{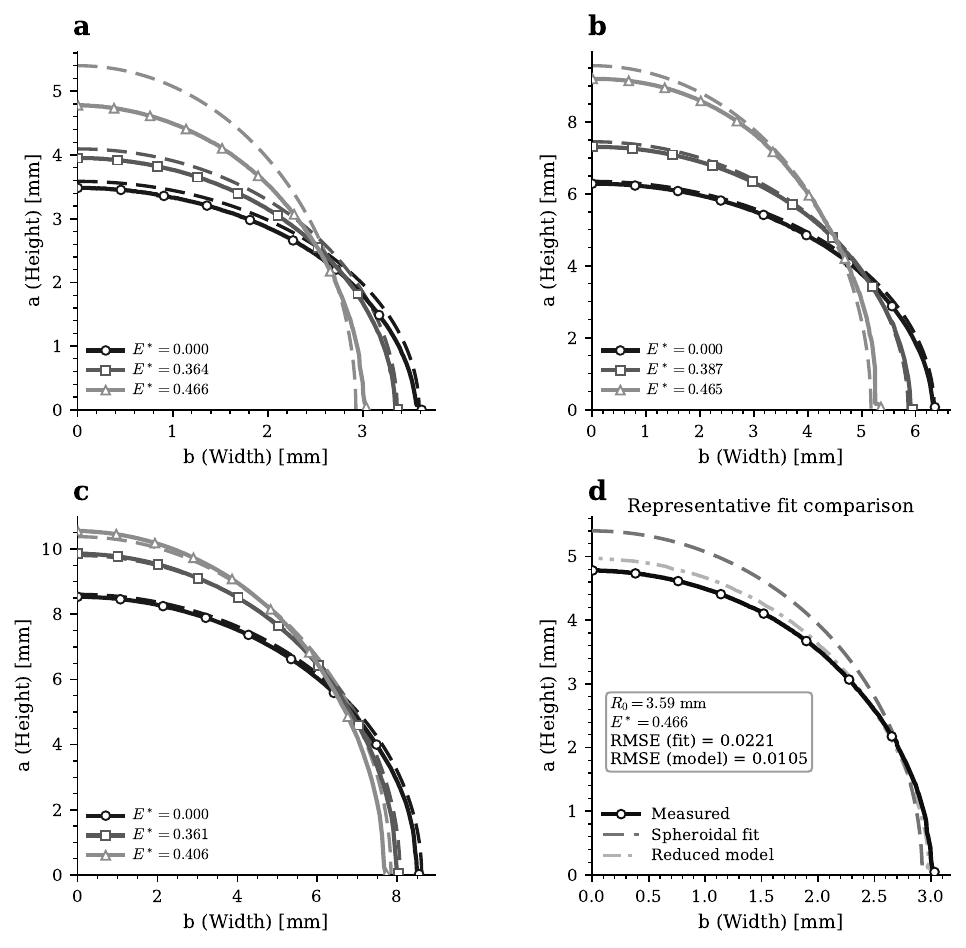}
\caption{\label{fig:stable_profiles}
Stable-regime profiles and spheroidal representation. Panels (a)--(c) show representative small-, intermediate-, and large-bubble meridional profiles for $R_0 = 3.59$, $6.35$, and $8.61$ mm at several electric-field levels. Solid lines with markers denote measured profiles and dashed lines denote spheroidal fits used to extract $\alpha = a/b$. Panel (d) shows a representative high-field stable small-bubble case together with a secondary two-mode empirical reconstruction discussed in Appendix~\ref{app:two_mode}.}
\end{figure*}

In the stable regime, the bubble relaxes after each voltage increment to a smooth steady shape. Fig.~\ref{fig:stable_profiles} shows representative meridional profiles for three initial-size groups. Over the measured electric-field range, the steady profiles along the branch are well represented by a spheroidal fit, which provides both a one-parameter fit and a consistent deformation metric through $\alpha = a/b$. As noted in Section II.B, the spheroidal fit serves as a deformation metric. Appendix A compares the spheroidal fit with a two-mode empirical reconstruction referenced to the zero-field profile of the same bubble. Because the reduction in median normalized-profile RMSE is modest (0.0135 to 0.0102), we retain the spheroidal fit and the aspect ratio $\alpha$ as the primary deformation metric.

\subsection{Collapse onto a single steady-state branch and transition to instability}

\begin{figure*}[t]
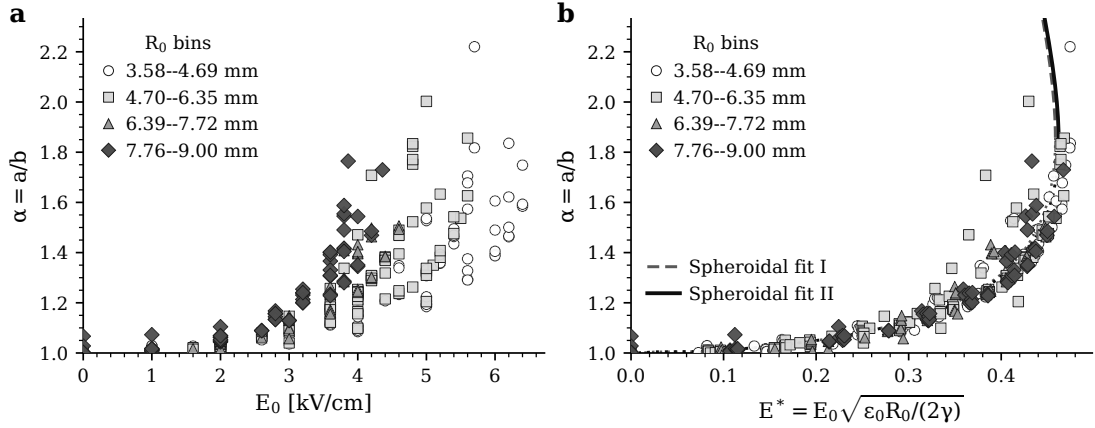

\centering
\begin{minipage}[t]{0.9\textwidth}
    \centering
    {\small\textbf{}}\\[2pt]
    \maybeincludegraphics[width=\textwidth]{figures/fig4_ab_combined.pdf}
\end{minipage}
\hfill
\caption{\label{fig:collapse} (a) Stable-regime aspect ratio plotted against the nominal applied field $E_0$. The systematic offset among $R_0$ bins shows that the raw deformation data remain size dependent in $E_0$. (b) Stable regime collapse onto a single steady-state branch by replotting the same data against $E^\ast$. Aspect ratio $\alpha$ versus dimensionless field $E^\ast$ collapses across multiple initial radii for data acquired within a single experimental session. The last observable steady states mark the endpoint of the branch and the transition to the unstable regime. The solid and dashed curves show two spheroidal baselines recast in the present normalization.
}
\end{figure*}

When $\alpha$ is plotted against $E^\ast$, data from multiple initial radii merge into a single steady-state branch within the tested range (Fig.~\ref{fig:collapse}). Over the tested size range and within a fixed ambient session, the deformation is organized primarily by electric stress relative to capillary pressure rather than by initial size alone. 

The endpoint of the branch is identified by slowly ramping the voltage near onset and recording the point at which the apex ceases to relax to a smooth steady profile and instead sharpens continuously. We denote the corresponding transition field by $E_c^\ast$.
Because $\alpha$ is single-valued on the steady-state branch, the same transition can equivalently be characterized by a corresponding aspect ratio $\alpha_c=\alpha(E_c^\ast)$. Over the tested size groups, the last observable steady states yield
$E^{*}_{c} \approx 0.45$--$0.47$ and $\alpha_c \approx 1.8$--$1.9$.
The narrow spread in $E^{*}_{c}$ indicates that the branch endpoint depends
only weakly on initial radius. The curve is reported for fixed ambient conditions in the present geometry. Control runs show that humidity shifts the branch between sessions, so $E^\ast$ should be read here as the organizing dimensionless field for the present system under fixed ambient conditions, not a session-independent constitutive law.

\subsection{Interpretive role of the spheroidal baseline for the steady-state branch}

The full electrohydrodynamic free-boundary problem stated above is not solved exactly in the present finite-gap, sessile, and thin-film geometry. Instead, we use a Taylor-type spheroidal approximation as a reduced baseline for the stable regime. Here, ‘Fit I’ denotes the spheroidal approximation obtained from a local balance at the pole and equator, whereas ‘Fit II’ denotes the companion approximation based on an integrated balance over the interface. 

That baseline gives a one-parameter description of the steady-state branch and reduces the stable data to $\alpha(E^\ast)$. In many spheroidal models, the steady branch terminates at a limiting electric loading~\cite{Taylor1964,TorzaCoxMason1971,BasaranScriven1989,BasaranScriven1990}. The experimentally observed last steady states qualitatively match such an endpoint, but the primary experimental result here is the collapsed branch itself, with the endpoint extracted from it.

\section{Cone formation and pre-jet apex dynamics}

\subsection{Cone-angle selection in a noncanonical geometry}

\begin{figure*}[t]
\centering

\begin{minipage}[t]{0.44\textwidth}
    \centering
    \maybeincludegraphics[width=\textwidth]{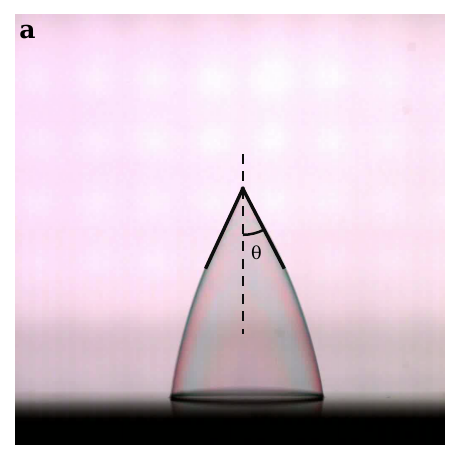}
\end{minipage}
\hfill
\begin{minipage}[t]{0.54\textwidth}
    \centering
    \maybeincludegraphics[width=\textwidth]{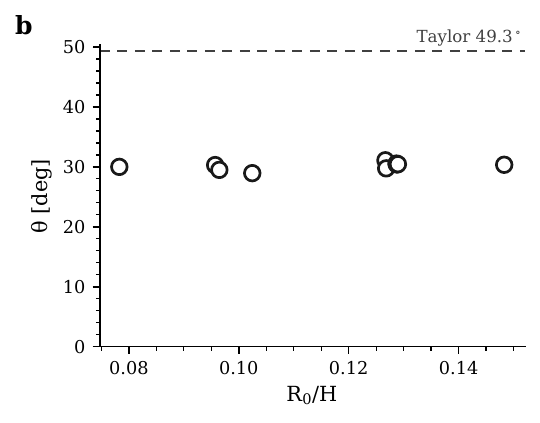}
\end{minipage}

\caption{\label{fig:fig5}
Cone-angle selection in the unstable regime. (a) Definition of the cone half-angle $\theta$ from local linear fits to the near-tip sides in the last frame before jetting. (b) Measured cone half-angle $\theta$ plotted against $R_0/H$ (equivalently, against the initial bubble size for fixed $H$). The dashed horizontal line marks the classical Taylor value $49.3^{\circ}$ only as an external reference for the ideal conical equipotential construction.}
\end{figure*}

Beyond the endpoint of the steady-state branch, or equivalently for $E^\ast>E_c^\ast$, the apex no longer approaches a steady smooth shape but sharpens rapidly into a pointed tip that is locally well represented by a cone over a finite near-tip region. The measured cone half-angle lies in the range $\theta \approx$ $30.0^{\circ}$ $\pm$ $0.6^{\circ}$, substantially below the classical Taylor value of $49.3^{\circ}$ (Fig.~\ref{fig:fig5}(b)). The reported angle range reflects both variation between experiments and the fact that the measured angle depends somewhat on which visible part of the cone side was used for the line fit. The cone angle is reproducibly selected in the sessile-bubble geometry under parallel-plate electric field with weak dependence on $R_0/H$.

This result does not contradict Taylor's conical construction~\cite{Taylor1964}. Taylor's $49.3^{\circ}$ half-angle follows from a specific electrostatic construction for an equipotential conical surface in a compatible field, and self-similar analyses of cone formation on conducting liquids recover this angle as the asymptotic attractor \cite{Zubarev2001}. The present experiments instead involve a parallel plate electrode geometry, a sessile bubble pinned on a substrate, and a soap film that need not behave as a perfectly conducting equipotential surface at all times. The present measurements are therefore more naturally interpreted as evidence of noncanonical angle selection, consistent with prior work on finite-geometry electrified drops and non-Taylor critical shapes~\cite{StoneListerBrenner1999,Yarin2001}.

\subsection{Apex approach before jetting}

\begin{figure*}[t]
\centering

\begin{minipage}[t]{0.5\textwidth}
    \centering
    \maybeincludegraphics[width=\textwidth]{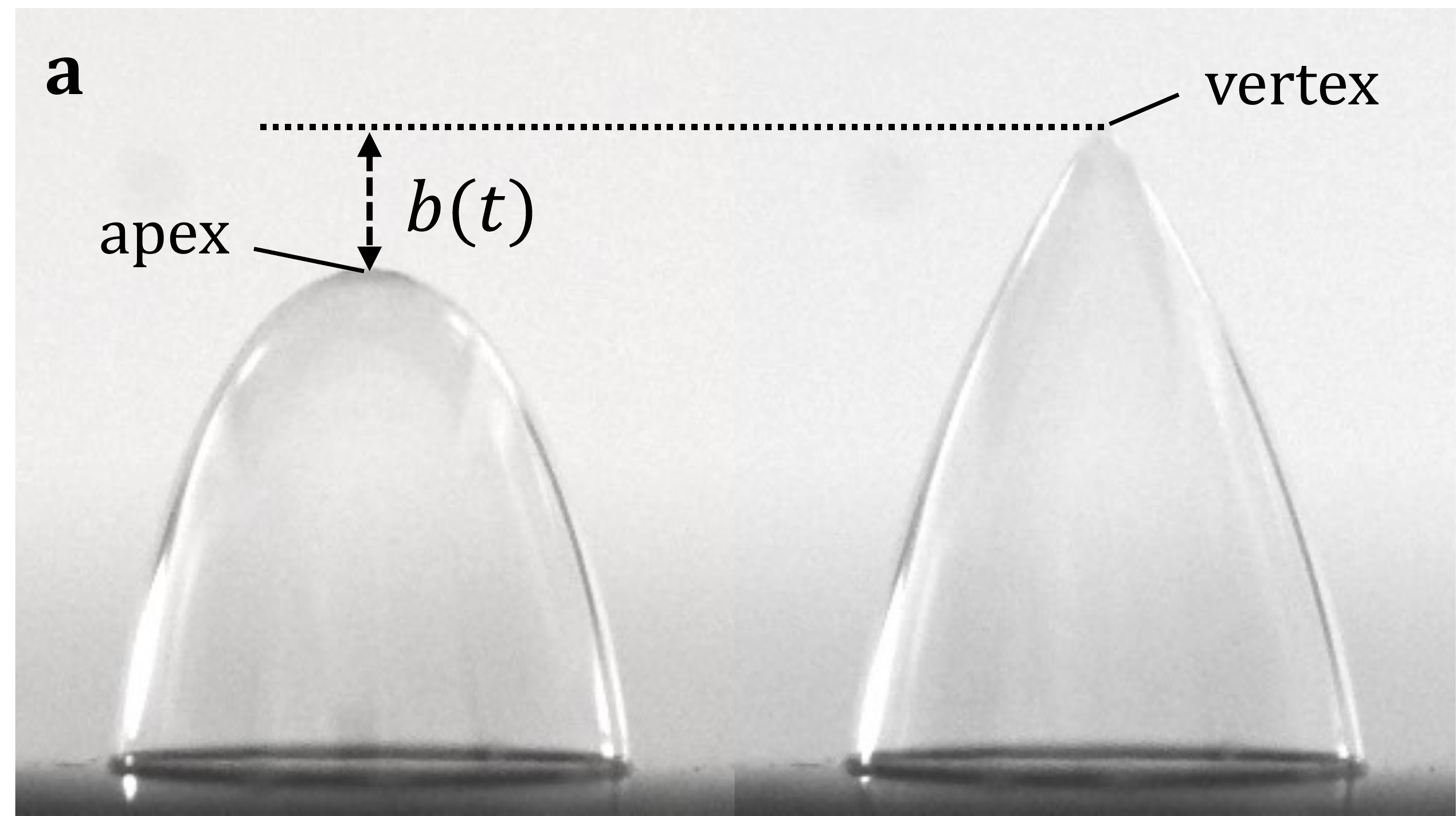}
\end{minipage}
\vfill
\begin{minipage}[t]{0.9\textwidth}
    \centering
    \maybeincludegraphics[width=\textwidth]{figures/fig6bc_v11.pdf}
\end{minipage}

\caption{\label{fig:fig6}
Unstable kinematics at the pre-jet stage. (a) Definition of the fixed reference vertex and the apex-to-reference-vertex distance $b(t)$. The reference vertex is determined from the first frame with visible jet emission by extrapolating the near-tip. The distance $b(t)$ for earlier frames is then measured from the instantaneous apex to this fixed reference point. (b) Temporal evolution of the apex-to-reference-vertex distance $b(t)$ with time aligned so that $t = 0$ marks the first frame with visible jet emission. (c) Comparison between measured 
$b$ and fitted $b$ for the points included in the fit in panel (b). The diagonal line indicates perfect agreement. 
}
\end{figure*}

To quantify the final sharpening stage, we define a fixed reference vertex from the first frame with visible jet emission, or equivalently from the last pre-jet frame for which the cone side profiles are stationary. Specifically, local linear fits are applied to the near-tip conical profile in that reference frame, and their extrapolated intersection defines the reference vertex, as shown in  Fig.~\ref{fig:fig6}(a). For each earlier frame, $b(t)$ is the axial distance from the instantaneous apex to this fixed reference vertex. With this definition, $b(t)$ decreases rapidly as jetting is approached, while the local curvature scale grows approximately as $\kappa \sim 1/b$. Time is aligned so that $t = 0$ marks the first frame with visible jet emission. The point closest to visible jet emission is the most sensitive to the choice of reference frame and to the line fit of the cone side, so it is excluded from the direct fit in Fig.~\ref{fig:fig6}(b) and from the parity plot comparing data with the model in Fig.~\ref{fig:fig6}(c).

The data in Fig.~\ref{fig:fig6}(b) show that $b$ decreases rapidly during the final measured interval before visible jet emission. To interpret that trend, we use a local force balance adapted from Taylor's original calculation of the motion of a conical soap-film interface toward its vertex~\cite{Taylor1964}, together with ideas from capillary--inertial singularity dynamics \cite{Eggers1997,EggersFontelos2015} and self-similar analyses of electrified tip formation \cite{Zubarev2001}. In this balance, the film carries an effective areal mass $m \sim \rho h$, and the restoring capillary force scales with curvature as $1/b$. This balance is used here only to test whether the measured decrease of $b$ is consistent with the logarithmic form implied by Eq.~(\ref{eq:blog}) when the comparison is performed directly in the time domain.

A local force balance then takes the form
\begin{equation}
\rho h\,\frac{d^2 b}{dt^2} \simeq -K\frac{\gamma}{b},
\label{eq:bddot}
\end{equation}
where $K$ is a dimensionless coefficient that collects the near-tip electric contribution into an effective prefactor. Multiplying Eq.~(\ref{eq:bddot}) by $db/dt$ and integrating gives
\begin{equation}
\left|\frac{db}{dt}\right| \sim \left(\frac{2K\gamma}{\rho h}\right)^{1/2}\left[\ln\left(\frac{b_0}{b}\right)\right]^{1/2},
\label{eq:blog}
\end{equation}
where $b_0$ is an outer cutoff of order $R_0$. Eq.~(\ref{eq:blog}) predicts a weak logarithmic increase in the rate of decrease of $b$ as $b \to 0$. Since the local near-tip field is not known, we do not interpret the prefactor quantitatively. Instead, we compare the measured $b(t)$ trajectories with the integrated form of Eq.~(\ref{eq:blog}). Fig.~\ref{fig:fig6}(c) shows the corresponding comparison between the measured data and the prediction for the points used in the fit.

\section{Discussion}

The stable-regime measurements show that the primary experimental result is the collapse of the aspect ratio onto a single steady-state branch $\alpha(E^\ast)$ across the tested initial radii within a fixed ambient session. In that sense, the transition field is not an independent parameter but the endpoint of the observed branch. Beyond that endpoint, the system enters a regime of conical sharpening and rapid apex approach before jetting.

For conducting sessile drops in parallel-plate fields, Basaran and Scriven~\cite{BasaranScriven1989,BasaranScriven1990} computed steady branches with turning points corresponding to terminal aspect ratios in the range $\alpha_c \approx 1.8$--$1.9$, depending on contact angle and electrical boundary conditions. The terminal aspect ratios measured here are close to $\alpha_c \approx 1.9$, despite the present interface being a soap film with two air--liquid surfaces rather than a bulk liquid drop. This agreement supports the use of the electric Bond number as a useful organizing parameter for the stable branch at the scaling level adopted here, while not implying that the present thin-film geometry is dynamically equivalent to the conducting-drop problem.

That qualification is important because the nondimensionalization uses the nominal background field $E_0 = V/H$, and the film thickness and electrical properties of the solution have not been measured directly. These limitations do not alter the empirical observations themselves, namely the collapsed steady-state branch, the operationally identified transition field, the measured cone-angle range, or the observed pre-jet trajectories in $b(t)$. They do, however, limit how far the present measurements can be pushed toward full mechanistic closure. In particular, a self-consistent calculation of the evolving near-tip electric field, together with in situ film-thickness measurement and direct characterization of conductivity, would be needed to connect the measured trajectories more tightly to a predictive electrohydrodynamic model.

The measured cone half-angle of $30.0^{\circ} \pm 0.6^{\circ}$ is substantially smaller than Taylor's $49.3^{\circ}$ and falls in the same broad range as the non-Taylor values reported in finite-geometry electrified-drop and jet systems~\cite{StoneListerBrenner1999,Yarin2001}. Yarin \emph{et al.}~\cite{Yarin2001} reported half-angles of about $33.5^{\circ}$ for polymer-solution jets in a parallel-capacitor configuration and interpreted the departure from $49.3^{\circ}$ in terms of boundary conditions that do not satisfy the self-similar assumptions behind Taylor's construction. The present measurements support the same general interpretation. In the sessile-bubble geometry, the interface is pinned to a substrate, the imposed field is set by a finite plate gap, and the soap film need not behave as a perfectly conducting equipotential surface throughout the sharpening process. The observed angle is therefore better understood as noncanonical angle selection in the present geometry than as a departure from a value expected to hold here. Whether the remaining difference between about $30^{\circ}$ here and about $33.5^{\circ}$ in Ref.~\cite{Yarin2001} reflects thin-film structure, contact conditions, or different fluid properties remains open.

The humidity sensitivity documented in Appendix~\ref{app:controls} further shows that the steady-state branch is conditional on ambient conditions. In the control runs, lower ambient humidity shifts the deformation curve to higher nominal electric field. This trend is consistent with humidity-dependent evolution of the soap film during the voltage ramp, although the present measurements do not isolate the responsible mechanism. Possible contributors include evaporation-driven thinning, changes in charge-relaxation behavior, and changes of permittivity in the surrounding dielectric medium. Resolving their relative importance would require in situ thickness measurements and electrical characterization during the field ramp.

The same distinction between empirical organization and predictive closure applies to Eq.~(\ref{eq:blog}). The logarithmic form contains quantities such as $K$, $b_0$, and $h$ that are not independently determined here, so its value is diagnostic rather than predictive. What the comparison in Fig.~\ref{fig:fig6} establishes is that the measured $b(t)$ trajectories are consistent with a reduced near-tip inertia-capillary balance at a shrinking tip. Eq.~(\ref{eq:blog}) should therefore be read as a reduced law for the measured pre-jet trajectories, not as a predictive cone-jet theory. A predictive model would require a self-consistent treatment of the evolving near-tip electric field coupled to thin-film charge transport.

\section{Conclusions}

We studied sessile soap bubbles driven by a nominally uniform electric field and identified two reproducible response regimes within one experimental geometry. In the stable regime, the meridional profile is well represented by a spheroid, and the data reduce to a single curve when plotted against the dimensionless field $E^\ast = \sqrt{\mathrm{Bo}_e}$ for data acquired within a fixed ambient session. The endpoint of this branch defines the transition to the unstable regime. Beyond that transition, the interface develops a pointed tip and then a cone with half-angle $30.0^{\circ}$ $\pm$ $0.6^{\circ}$, below the classical Taylor value. A fixed reference vertex provides a practical reference for comparing the late stage before jetting across bubble sizes. The measured decrease of $b(t)$ accelerates as jetting is approached, and a reduced near-tip inertia-capillary model captures the observed logarithmic trend in that rate.

\begin{acknowledgments}
This work was supported by a graduate fellowship from SK Innovation Co., Ltd. 

The authors thank Crystal R. Fowler for her ideas and knowledge to develop this project. H.K. and S.J. conceived the idea, designed the experiment. H.K. built the experimental setup, collected and analyzed data, interpreted the results and wrote the manuscript. H.K. and S.J. revised the manuscript.

The authors have no competing interests.
\end{acknowledgments}

\FloatBarrier
\appendix

\section{Secondary stable-profile reconstruction}
\label{app:two_mode}

As a model competency check, we also compare the measured meridional profiles with a two-mode reconstruction using a data-driven method referenced to the zero field profile of the same bubble,
\begin{equation}
y(E^\ast) \approx y_0 + c_1(E^\ast)\psi_1 + c_2(E^\ast)\psi_2.
\end{equation}
Here $\psi_1$ and $\psi_2$ are empirical shape functions extracted from the stable-regime dataset, and the comparison is illustrated by Fig.~\ref{fig:stable_profiles}(d). Over stable cases with the field on, this representation reduces the median normalized-profile RMSE from $0.0135$ to $0.0102$. Because the improvement is modest and because the coefficients $c_1(E^\ast)$ and $c_2(E^\ast)$ are purely empirical, we retain the spheroidal fit and the aspect ratio $\alpha$ as the primary deformation metric in the main text.

\section{Operational definitions and humidity sensitivity}
\label{app:controls}

Near the transition to the unstable regime, the voltage was increased slowly from the highest clearly stable condition and the onset of persistent sharpening was recorded directly. We take that onset value as $E_c^\ast$, with uncertainty set by the voltage readout resolution near onset together with the finite sensitivity in identifying the first persistent departure from steady relaxation. In the main text, this onset is interpreted as the endpoint of the steady-state branch in Fig.~\ref{fig:collapse}.

A similar operational issue enters the unstable regime metric. The fixed reference vertex is determined from the first frame with visible jet emission, or equivalently from the last pre-jet frame for which the cone shapes are stationary. The data point nearest jet emergence is excluded from the rate comparison because it is the most sensitive to the selected reference frame, finite frame rate, and the line fit to the cone side.

Separate control experiments show that lower ambient humidity shifts the deformation curve to higher nominal electric. Because several coupled mechanisms could contribute, including changes in permittivity of the medium which is air in this case, charge relaxation, and evaporation-driven film evolution, humidity is treated here as an experimental control variable.

\begin{figure*}[t]
\centering
\maybeincludegraphics[width=0.85\textwidth]{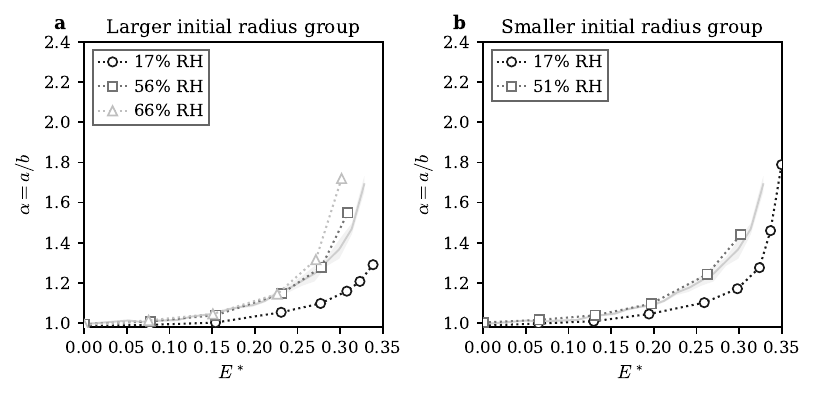}
\caption{Humidity-control measurements for the stable regime. Panel (a) shows the larger initial-radius group and panel (b) shows the smaller initial-radius group. The shift of the deformation curve with relative humidity indicates that the collapsed steady-state branch in Fig.~\ref{fig:collapse} is conditional on the ambient condition of a given session. The thin gray band indicates the reference stable branch obtained from the main steady-state dataset and is shown only as a background for comparison.
}
\label{fig:humidity_control}
\end{figure*}

\begin{figure*}[t]
\centering
\maybeincludegraphics[width=0.9\textwidth]{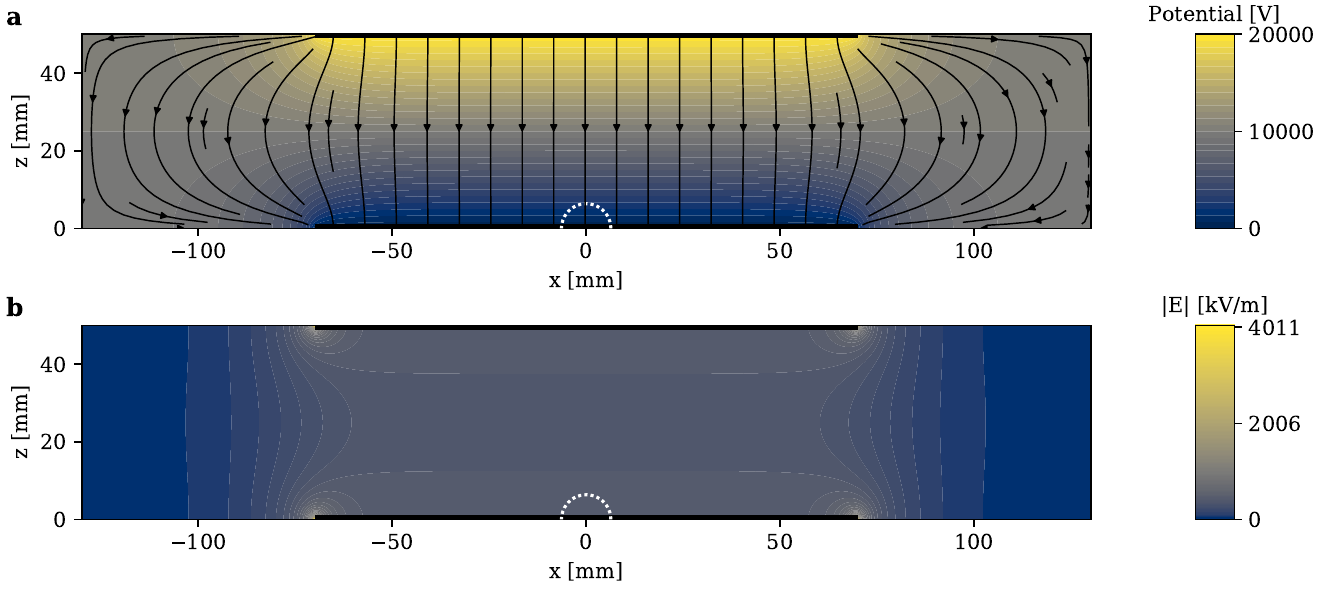}
\caption{Finite-electrode Laplace calculation for the experimental geometry.
(a) Electric potential and representative field lines.
(b) Electric-field magnitude.
The dashed semicircle marks a representative bubble-scale guide centered beneath the top plate. It is a geometric guide for post-processing only and is not included in the Laplace solve. Evaluated along this guide, the axial field differs from the parallel-plate estimate $V/H$ by less than about $0.02\%$, supporting the use of $E_0 = V/H$ as the nominal field in the main-text stable-regime scaling.}
\label{fig:fieldsim}
\end{figure*}

\section{Finite-electrode Laplace calculation for the nominal background field}
\label{app:theta_scatter}

To assess how closely the experimental geometry approximates a parallel-plate field, we solved Laplace's equation for the finite electrode configuration used in the experiment. The purpose of this calculation is limited to the nominal background field over the bubble-scale region beneath the plate center. It is not used to estimate the local field in the apex region during conical sharpening or jetting.

Fig.~\ref{fig:fieldsim}(a) shows the electric potential and representative field lines for the imposed finite-electrode geometry, and Fig.~\ref{fig:fieldsim}(b) shows the corresponding electric-field magnitude. Because the quantity relevant to the nondimensionalization is the imposed axial background field, the comparison with the nominal estimate $V/H$ is made from the axial component extracted from the same Laplace solution along the representative bubble-scale guide indicated in Fig.~\ref{fig:fieldsim}.

Over that bubble scale region, the axial field differs from the parallel-plate estimate $V/H$ by less than about $0.02\%$, with a mean deviation of about $+0.02\%$. This supports the use of $E_0 = V/H$ as the nominal applied field in the stable-regime scaling. The dashed semicircle in Fig.~\ref{fig:fieldsim} is a geometric guide for this post-processing comparison only and is not included in the Laplace solve.

\FloatBarrier
\bibliographystyle{apsrev4-2}
\bibliography{references_1_revised}

\end{document}